# Supersymmetric Yang-Mills Theory in Eleven Dimensions[1]

Hitoshi NISHINO[2]

*Department of Physics*
*University of Maryland at College Park*
*College Park, MD 20742-4111, USA*

## Abstract

We present a Lorentz invariant lagrangian formulation for a supersymmetric Yang-Mills vector multiplet in eleven dimensions (11D). The Lorentz symmetry is broken at the field equation level, and therefore the breaking is spontaneous, as in other formulations of supersymmetric theories in 12D or higher dimensions. We introduce a space-like unit vector formed by the gradient of a scalar field, avoiding the problem of Lorentz non-invariance at the lagrangian level, which is also an analog of non-commutative geometry with constant field strengths breaking Lorentz covariance. The constancy of the space-like unit vector field is implied by the field equation of a multiplier field. The field equations for the physical fields are formally the same as those of 10D supersymmetric Yang-Mills multiplet, but now with some constraints on these fields for supersymmetric consistency. This formulation also utilizes the multiplier fields accompanied by the bilinear forms of constraints, such that these multiplier fields will not interfere with the physical field equations. Based on this component result, we also present a $\kappa$-symmetric supermembrane action with the supersymmetric Yang-Mills backgrounds.

---

[1]This work is supported in part by NSF grant # PHY-93-41926.
[2]E-Mail: nishino@nscpmail.physics.umd.edu

## 1. Introduction

It has been a common wisdom that the maximal space-time dimensions for an interacting vector multiplet are ten dimensions (10D), which can accommodate a supermultiplet with the maximal helicity one [1][2], as long as we keep Lorentz invariance manifest, unlike those formulations in [3][4] which introduce constant null-vectors. The traditional argument for the obstruction to go beyond 10D for a vector multiplet is that the number of total on-shell bosonic and fermionic physical degrees of freedom for a Maxwell vector multiplet in 10D is $8+8$, while in 11D it becomes $9+16$ with the $16$ fermionic degrees of freedom exceeding the $9$ bosonic ones. This mismatch problem becomes worse in dimensions $D \geq 12$, because the number of fermionic degrees of freedom for a Majorana spinor $\lambda$ grows like $2^{[D/2]-1}$, while that of a vector field $A_\mu$ grows only linearly like $D-2$. An independent conclusion for 10D as the maximal space-time dimensions is based on the reformulation of vector multiplets in terms of octonions (Cayley numbers) [1], because the octonion field forms the maximal division algebra, according to the theorem of Hurwitz [5][1]. Considering these independent 'no-go' theorems, we see no hope of formulating any supersymmetric vector multiplet in $D \geq 11$ with a Lorentz invariant Lagrangian, and any 'ignorant' trial breaking the traditional 'taboo' seems to be doomed to fail.

However, we point out that one important ingredient has been routinely overlooked in the traditional arguments above, namely we may still be able to present certain Lorentz invariant lagrangian, even though the field equations may lack Lorentz covariance. In other words, the Lorentz covariance may be broken 'spontaneously' at the field equation level, while the lagrangian maintains the Lorentz invariance. As a matter of fact, such phenomena have been common in non-commutative geometry in string theories [6], in which certain field strengths get constant values breaking Lorentz covariance. Such a formulation with broken Lorentz invariance was originated by Dirac [7], and was also independently developed by Kostelecký and Samuel [8] motivated by bosonic string theory in higher dimensions.

The trial of formulating such a supersymmetric vector multiplet in 11D is also strongly motivated by M-theory [9][10] because D-branes [11] play crucial roles for studying the non-perturbative aspects of M-theory, where the Born-Infeld action describes the $U(1)$ gauge field on the D-brane worldvolume. The success of the BFSS conjecture [9] of reducing the 10D supersymmetric Yang-Mills lagrangian dimensionally either into 1D or 0D, used as a matrix model generating effective potentials for 11D supergravity [9][10], also indicates a possible direct formulation of a vector multiplet within 11D. Therefore, it is an imperative to ask whether such a formulation of a vector multiplet is possible in 11D from the outset.

In this paper, we present a Lorentz invariant lagrangian with a supersymmetrically invariant action within 11D. Our formulation is based on the combination of two important recent techniques: First, we introduce a space-like 'harmonic' unit vector $B_\mu$ satisfying $B_\mu^2 = 1$ [12] which is non-constant from the outset, but is forced to be constant only on-shell. Second, we use of Siegel's multiplier fields [13], avoiding the problem of multipliers involved in 'physical' field equations. The introduction of such an 'on-shell' constant vector is also similar to the above-mentioned non-commutative geometry in string theory [6] with constant field strengths. Subsequently, we confirm the three fundamental consistencies of the system, *i.e.,* the consistency among field equations, the invariance of the action under supersymmetry, and the closure of supersymmetry on all the fundamental fields in our theory. Based on the component result, we present a $\kappa$-invariant supermembrane action with a Chern-Simons-Hopf term coupled to the 11D supersymmetric Yang-Mills backgrounds.

## 2. Total Lagrangian and Field Equations in 11D

We start with our total lagrangian in 11D, and study the field equations of all the fields. Our fundamental field content is $(A_\mu{}^I, \lambda_\alpha{}^I, \varphi, \Lambda_{(2)}^{\mu\nu}, \Xi^{\mu_1\cdots\mu_r(i),\,\nu_1\cdots\nu_s(j)})$, where $A_\mu{}^I$ is the usual potential field for the Yang-Mills multiplet with the adjoint index $I, J, \cdots$, whose field strength is $F_{\mu\nu}{}^I \equiv \partial_\mu A_\nu{}^I - \partial_\nu A_\mu{}^I + f^{IJK} A_\mu{}^J A_\nu{}^K$, $\lambda_\alpha{}^I$ is the gaugino, $\varphi$ is a real scalar field, $\Lambda_{(2)}^{\mu\nu}$ is a multiplier filed, and $\Xi^{\mu_1\cdots\mu_r(i),\,\nu_1\cdots\nu_s(j)}$ $(r,\,s = 0,\,1,\,\cdots;\,i,\,j = 1,\,\cdots,\,4)$ are also multiplier fields, whose role will be clarified shortly.

Our total action $S$ is composed of three lagrangians $\mathcal{L}_0$, $\mathcal{L}_\Lambda$ and $\mathcal{L}_\Xi$:[3]

$$S \equiv \int d^{11}x\, \mathcal{L} \equiv \int d^{11}x\, (\mathcal{L}_0 + \mathcal{L}_\Lambda + \mathcal{L}_\Xi) \ , \tag{2.1a}$$

$$\mathcal{L}_0 \equiv -\tfrac{1}{4}(F_{\mu\nu}{}^I)^2 + (\overline{\lambda}^I \gamma^\mu D_\mu \lambda^I) \ , \tag{2.1b}$$

$$\mathcal{L}_\Lambda \equiv \Lambda_{(2)}^{\mu\nu} C_{(2)\mu\nu} \ , \tag{2.1c}$$

$$\mathcal{L}_\Xi \equiv \tfrac{1}{2} \sum_{r=0}^{\infty} \sum_{s=0}^{\infty} \Xi^{\mu_1\cdots\mu_r(i),\,\nu_1\cdots\nu_s(j)} C_{\nu_1\cdots\nu_s(j)} C_{\mu_1\cdots\mu_r(i)} \ . \tag{2.1d}$$

Thus $\mathcal{L}_0$ is formally the same as the 10D case with the Yang-Mills covariant derivative $D_\mu$, while $\mathcal{L}_\Lambda$ is a multiplier lagrangian and $\mathcal{L}_\Xi$ is based on the formulation originally developed by Siegel [13] using the bilinear form of constraints. Even though there is formally an infinite number of multiplier fields $\Xi$, these terms should be terminated as a finite sum of terms, due to the finiteness requirement of the total action [14][15]. We will come back to this point at a later stage. The $C$'s in (2.1d) for $r = s = 0$ are the following combinations of our fundamental fields[4]

$$C_{(1)\alpha I} \equiv (N\lambda^I)_\alpha \equiv \tfrac{1}{2}(I - \beta)_\alpha{}^\beta \lambda_\beta{}^I \equiv \tfrac{1}{2}(I - B_\mu \gamma^\mu)_\alpha{}^\beta \lambda_\beta{}^I \ , \tag{2.2a}$$

$$C_{(2)\mu\nu} \equiv \partial_\mu B_\nu \equiv \partial_\mu \left[ \frac{\partial_\nu \varphi}{\sqrt{(\partial_\rho \varphi)^2}} \right] \ , \tag{2.2b}$$

$$C_{(3)\mu I} \equiv B^\nu F_{\mu\nu}{}^I \ , \tag{2.2c}$$

$$C_{(4)\alpha I} \equiv B^\mu (D_\mu \lambda_\alpha{}^I) \ , \tag{2.2d}$$

where the indices *after* $(i)$ will be sometimes omitted to make the expression more compact. The $B_\mu$ and $\beta$ are defined in terms of $\varphi$ by

$$B_\mu \equiv \frac{\partial_\mu \varphi}{\sqrt{(\partial_\nu \varphi)^2}} \ , \quad \beta \equiv B_\mu \gamma^\mu \ , \tag{2.3}$$

---

[3] We use the signature $(\eta_{\mu\nu}) = \text{diag.}\,(-,+,+,\cdots,+)$ $(\mu,\,\nu,\,\cdots = 0,\,1,\,\cdots,\,10)$ in this paper.

[4] We use the indices $I, J, \cdots$ on the l.h.s. always as *subscripts* for the reason to be clarified shortly. As long as the Yang-Mills gauge group has the positive definite metric $\delta^{IJ}$, such a mismatch of raisings/lowerings of these indices will not matter.

with the relevant properties

$$B_\mu^2 \equiv +1 \ , \quad \beta^2 \equiv +I \ . \tag{2.4}$$

This formulation using a scalar auxiliary field to express a space-like 'harmonic' unit vector $B_\mu$ was first presented in [12] for duality-symmetric actions. We utilize this technique here in order to avoid the introduction of 'constant' vectors, that explicitly break the Lorentz invariance of the total lagrangian.

The matrix $\beta$ is an analog of $\gamma_{11}$ in 10D. Accordingly, the matrix $N$ is a 'negative chirality projector' defined also with the positive counter part $P$ as

$$P \equiv \tfrac{1}{2}(I + \beta) \ , \quad N \equiv \tfrac{1}{2}(I - \beta) \ , \tag{2.5a}$$

$$P^2 \equiv P \ , \quad N^2 \equiv N \ , \quad PN \equiv NP \equiv 0 \ , \tag{2.5b}$$

$$\beta P \equiv P\beta \equiv +P \ , \quad \beta N \equiv N\beta \equiv -N \ . \tag{2.5c}$$

Even though we named them 'projectors', they still maintain the Lorentz covariance within 11D, as long as the constraint $C_{(2)\mu\nu} = 0$ is *not* imposed.

The $C_{\mu_1\cdots\mu_r(i)}$ in (2.1d) is defined as the $r$-th covariant derivative of $C_{(i)}$ by

$$C_{\mu_1\cdots\mu_r(i)} \equiv D_{\mu_1} D_{\mu_2} \cdots D_{\mu_r} C_{(i)} \ . \tag{2.6}$$

Therefore, the vectorial/spinorial indices *after* the indices $(i), (j), \cdots$ are for the indices associated with the original $C_{(i)}$'s in (2.2), while the vectorial indices *before* $(i), (j), \cdots$ are for these covariant derivatives, to be distinguished from the former ones. For example, $C_{\mu(1)\alpha I} \equiv D_\mu C_{(1)\alpha I}$, where the indices $\alpha I$ are written explicitly.

We now study all the field equations for our fields. First, the field equations of the multiplier $\Lambda_{(2)}^{\mu\nu}$ yields simply

$$C_{(2)\mu\nu} \equiv \partial_\mu B_\nu \equiv \partial_\mu \left[ \frac{\partial_\nu \varphi}{\sqrt{(\partial_\rho \varphi)^2}} \right] \doteq 0 \ , \tag{2.7}$$

where the symbol $\doteq$ is for a field equation to be distinguished from other simple equations or algebraic identities. Once this field equation is imposed on $\varphi$, then the vector $B_\mu$ no depends on the coordinates, and thus it becomes a constant vector. Since this $B_\mu$ is also space-like, the most convenient choice of our coordinates is such that only $B_{11}$ is unity, while all other components are zero. Accordingly, we have the solution for $\varphi$ as

$$\varphi \doteq a_\mu x^\mu + c = a_{10} x^{10} + c \ ,$$
$$B_\mu \doteq \frac{a_\mu}{\sqrt{a_\nu^2}} \doteq \begin{cases} 1 & (\text{for } \mu = 10) \ , \\ 0 & (\text{for } \mu \neq 10) \ , \end{cases} \tag{2.8}$$

where $a_{10} \neq 0$ and $c$ are real constants. By this field equation, the system loses the 11D Lorentz covariance 'on-shell', and it is reduced into a 10D system.

We point out that the introduction of 'constant' vector in terms of a scalar field is similar to the recently-discovered non-commutative geometry [6] in string theory. This is because in non-commutative geometry [6], it is common to have constant field strengths that break Lorentz covariance. Our $B_\mu$ is proportional to a 1-form field strength $\partial_\mu\varphi$ for the 'potential' $\varphi$ that develops constant v.e.v., breaking the Lorentz covariance from the original $SO(10,1)$ in 11D into $SO(9,1)$ in 10D.

The field equations for the $\Xi$-fields need special care. First, we note that these fields for $r = s = 0$ have the (anti)symmetry with respect to their indices:

$$\Xi^{(i),(j)} = (-1)^{(i)(j)} \Xi^{(j),(i)} \quad , \tag{2.9}$$

where the indices $(i), (j), \cdots$ denote also their intrinsic tensorial/spinorial indices. For example, the index $(1)$ stands also for *hidden* indices $\alpha I$, as is clear from the constraint (2.2a). Relevantly, in (2.9) we use the symbol $(-1)^{(i)(j)}$ for the Grassmann parity for these two indices $(i)$ and $(j)$, including their *hidden* indices. Namely, if at least one of $(i)$ and $(j)$ is bosonic, then $(-1)^{(i)(j)} = +1$, while if both of them are fermionic, then $(-1)^{(i)(j)} = -1$. In other words, $\Xi$ is (anti)symmetric, *e.g.*, $\Xi^{(1)\alpha I,(4)\beta J} = -\Xi^{(4)\beta j,(1)\alpha I}$. Because of this (anti)symmetry, all the 'diagonal' components $\Xi^{(1),(1)}$, $\Xi^{(2),(2)}$, $\Xi^{(3),(3)}$, $\Xi^{(4),(4)}$ for $r = s = 0$ yield the constraints[5]

$$C_{(1)\alpha I} C_{(1)\beta J} \doteq 0 \quad \implies \quad C_{(1)\alpha I} \equiv (N\lambda^I)_\alpha \doteq 0 \quad , \tag{2.10a}$$

$$C_{(2)\mu\nu} C_{(2)\rho\sigma} \doteq 0 \quad \implies \quad C_{(2)\mu\nu} \equiv \partial_\mu B_\nu \equiv \partial_\mu \left[\frac{\partial_\nu\varphi}{\sqrt{(\partial_\rho\varphi)^2}}\right] \doteq 0 \quad , \tag{2.10b}$$

$$C_{(3)\mu I} C_{(3)\nu J} \doteq 0 \quad \implies \quad C_{(3)\mu I} \equiv B^\rho F_{\mu\rho}{}^I \doteq 0 \quad , \tag{2.10c}$$

$$C_{(4)\alpha I} C_{(4)\beta J} \doteq 0 \quad \implies \quad C_{(4)\alpha I} \equiv B^\mu (D_\mu \lambda_\alpha{}^I) \doteq 0 \quad . \tag{2.10d}$$

Needless to say, (2.10b) is also consistent with the $\Lambda$-field equation (2.7). Eq. (2.10a) effectively projects $\lambda$ into the positive chiral component in 10D, while (2.10c) deletes the 11-th component in $F_{\mu\nu}$, and (2.10d) is needed as its supersymmetric partner. Once these field equations are implied, it is also automatic to have field equations like

$$C_{\mu_1\cdots\mu_r(i)} \equiv D_{\mu_1} D_{\mu_2} \cdots D_{\mu_r} C_{(i)} \doteq 0 \qquad (\text{for } \forall r \geq 0) \quad . \tag{2.11}$$

Therefore, all the other 'non-diagonal' components of $\Xi^{(i),(j)}$ at $r = s = 0$ as well as higher components $\Xi^{\mu_1\cdots\mu_r(i),\nu_1\cdots\nu_s(j)}$ for $r \geq 1$ or $s \geq 1$ yield the field equations consistent with (2.11), and therefore no new information will be obtained, as far as the field equations are concerned.

As for the scalar field $\varphi$, its field equation is

$$\frac{\delta\mathcal{L}}{\delta\varphi} = \left(\frac{\delta B^\nu}{\delta\varphi}\right)\left(\frac{\delta\mathcal{L}}{\delta B_\nu}\right) = +\partial_\nu\partial_\mu\Lambda_{(2)}^{\mu\nu} + \mathcal{O}(\varphi\Lambda) \doteq 0 \quad , \tag{2.12a}$$

$$\frac{\delta\mathcal{L}}{\delta B_\nu} = -\partial_\mu\Lambda_{(2)}^{\mu\nu} + \sum_{r=0}^\infty \sum_{s=0}^\infty \Xi^{\mu_1\cdots\mu_r(i),\nu_1\cdots\nu_s(j)} C_{\nu_1\cdots\nu_s(j)} \frac{\delta}{\delta\varphi} C_{\mu_1\cdots\mu_r(i)} \quad . \tag{2.12b}$$

---

[5]Eq. (2.10a) and (2.10d) are valid, as long as we maintain *linear* supersymmetry. However, there is subtlety with the possibility of *non-linear* realization of supersymmetry with some nilpotent numbers [16] which we will discuss in the concluding remarks. See also the paragraph next to (3.3).

Thanks to the bilinear structure with $C$'s in $\mathcal{L}_\Xi$, the second term in (2.12b) vanishes upon (2.11), so that only the first term (2.12b) is important. We study the consistency of this field equation with supersymmetry in the next section.

The field equation for $A_\mu{}^I$ reads

$$\frac{\delta\mathcal{L}}{\delta A_\mu{}^I} = -D_\nu F^{\mu\nu\,I} - f^{IJK}(\overline{\lambda}^J\gamma^\mu\lambda^K)$$
$$+ \sum_{r=0}^\infty\sum_{s=0}^\infty \Xi^{\rho_1\cdots\rho_r(i),\,\sigma_1\cdots\sigma_s(j)} C_{\sigma_1\cdots\sigma_s(j)} \frac{\delta}{\delta A_\mu{}^I} C_{\rho_1\cdots\rho_r(i)} \doteq 0 \ . \qquad (2.13)$$

Here again, the last line vanishes upon (2.11), so that eventually the $A_\mu{}^I$-field equation is formally the same as that in 10D only in the direction of $\mu = 0, 1, \cdots, 9$ with the 'chirality' $\lambda \doteq P\lambda$. Anyhow, we have to keep in mind that these fields undergo the constraints (2.10).

Similarly, the gaugino field equation is

$$\frac{\delta\mathcal{L}}{\delta\lambda^I} = +2\gamma^\mu D_\mu\lambda^I \doteq 0 \ , \qquad (2.14)$$

which is again formally the same as that in 10D, when the contributions from $\mathcal{L}_\Xi$ vanished upon the field equations (2.10), like the second line in (2.13).

## 3. Supersymmetric Invariance of Total Action

We now confirm the invariance of our total action for the total lagrangian $\mathcal{L}$ under supersymmetry. Our supersymmetry transformation rule is dictated by

$$\delta_Q A_\mu{}^I = -(\overline{\epsilon}N\gamma_\mu\lambda^I) \ , \qquad (3.1a)$$

$$\delta_Q \lambda^I = -\tfrac{1}{4}\gamma^{\mu\nu}P\epsilon F_{\mu\nu}{}^I \ , \qquad (3.1b)$$

$$\delta_Q \varphi = 0 \ , \qquad (3.1c)$$

$$\delta_Q \Lambda_{(2)}{}^{\mu\nu} = -\tfrac{1}{4}(\overline{\epsilon}N\gamma^\nu\gamma^{\rho\sigma}\gamma^\mu\lambda^I)F_{\rho\sigma}{}^I \ , \qquad (3.1d)$$

$$\delta_Q \Xi^{\mu_1\cdots\mu_r(i),\,\nu_1\cdots\nu_s(j)} = -\sum_{t=0}^\infty \Xi^{\mu_1\cdots\mu_r(i),\,\rho_1\cdots\rho_t(k)} M_{\rho_1\cdots\rho_t(k)}{}^{\nu_1\cdots\nu_s(j)}$$
$$-\sum_{t=0}^\infty \Xi^{\nu_1\cdots\nu_s(j),\,\rho_1\cdots\rho_t(k)} M_{\rho_1\cdots\rho_t(k)}{}^{\mu_1\cdots\mu_r(i)}$$
$$\text{(for other than } \delta_Q\Xi^{(1)\alpha I,\,(1)\beta J}) \ , \qquad (3.1e)$$

$$\delta_Q\Xi^{(1)\alpha I,\,(1)\beta J} = +2f^{IJK}\Big[(\overline{\epsilon}N\gamma_\mu P\lambda^K)(\gamma^\mu)^{\alpha\beta} + 2(\overline{\epsilon}N)^{(\alpha|}(P\lambda^K)^{|\beta)}\Big]$$
$$-\sum_{t=0}^\infty \Xi^{(1)\alpha I,\,\rho_1\cdots\rho_t(k)} M_{\rho_1\cdots\rho_t(k)}{}^{(1)\beta J} - \sum_{t=0}^\infty \Xi^{(1)\beta J,\,\rho_1\cdots\rho_t(k)} M_{\rho_1\cdots\rho_t(k)}{}^{(1)\alpha I} \ , \qquad (3.1f)$$

where $M$'s are defined by

$$\delta_Q C_{\mu_1\cdots\mu_r(i)} = \sum_{s=0}^\infty M_{\mu_1\cdots\mu_r(i)}{}^{\nu_1\cdots\nu_s(j)} C_{\nu_1\cdots\nu_s(j)} \ . \qquad (3.2)$$

In general we omit the $\sum$-symbol for the dummy contractions for the indices $(i), (j), (k), \cdots$, while the summations over $r, s, \cdots$ are given explicitly. These $M$'s can contain any fields in our system, such as $F_{\mu\nu}{}^I$, $\lambda_\alpha{}^i$, $\Lambda^{\mu\nu}_{(2)}$, *etc.* or even $C$'s themselves, but their explicit expressions for general $r$ and $s$ are not so crucial in our formulation.

As examples, we give some explicit forms for $M_{\mu_1\cdots\mu_r(i)}{}^{\nu_1\cdots\nu_s(j)}$ for the case of $r = 0$:

$$M_{(1)\alpha I}{}^{(3)\nu J} = -\tfrac{1}{2}\delta^{IJ}(\gamma^\nu P\epsilon)_\alpha \ , \quad M_{(1)\alpha I}{}^{(1)} = M_{(1)\alpha I}{}^{(2)} = M_{(1)\alpha I}{}^{(4)} = 0 \ , \tag{3.3a}$$

$$M_{(2)\mu\nu}{}^{(j)} = 0 \ , \quad (j = 1, 2, 3, 4) \ , \tag{3.3b}$$

$$M_{(3)\mu I}{}^{\nu(1)\alpha J} = \delta^{IJ}\delta_\mu{}^\nu \epsilon^\alpha \ , \quad M_{(3)\mu\nu}{}^{(3)} = 0 \ , \tag{3.3c}$$

$$M_{(3)\mu I}{}^{(2)\rho\sigma} = \tfrac{1}{2}\delta_\mu{}^\rho(\overline{\epsilon}N\gamma_\sigma\lambda^I) + \tfrac{1}{2}\delta_\mu{}^\rho(\overline{\epsilon}\gamma^\sigma\beta\lambda^I) - \tfrac{1}{2}B^\rho(\overline{\epsilon}\gamma^\sigma\gamma_\mu\lambda^I) \ , \tag{3.3d}$$

$$M_{(3)\mu I}{}^{(4)\alpha J} = \delta^{IJ}(\overline{\epsilon}N\gamma_\mu)^\alpha \ , \tag{3.3e}$$

$$M_{(4)\alpha I}{}^{(1)\beta J} = f^{IJK}\lambda_\alpha{}^K(\overline{\epsilon}N)^\beta \ , \quad M_{(4)\alpha I}{}^{(4)} = 0 \ , \tag{3.3f}$$

$$M_{(4)\alpha I}{}^{(2)\mu\nu} = -\tfrac{1}{2}(\gamma^{\mu\rho}P\epsilon)_\alpha F_\rho{}^{\nu I} - \tfrac{1}{8}B^\mu(\gamma^{\rho\sigma}\gamma^\nu\epsilon)_\alpha F_{\rho\sigma}{}^I \ , \tag{3.3g}$$

$$M_{(4)\alpha I}{}^{\rho(3)\mu J} = \tfrac{1}{2}\delta^{IJ}(\gamma^{\rho\mu}P\epsilon)_\alpha \ , \tag{3.3h}$$

These are obtained by explicit application of the supersymmetry transformation to the $C$'s under question. With these explicit forms for $r = 0$, as well as using reductions with respect to the number of derivatives $r$, we can also conclude the validity of the general form (3.2) to all orders in $r$. Even though we do not give here the detailed all-order proof which is straightforward by induction with respect to $r$, it is important to note that we need the infinite sum in $\mathcal{L}_\Xi$ for the total action to be invariant under supersymmetry.

Another important conclusion from (3.3) is the consistency of all the constraint equations in (2.10) with supersymmetry. In particular, the previously-mentioned subtlety related to (2.10a) and (2.10d) is also resolved by (3.3). For example, once $C_{(3)}$ in (2.10c) vanishes, then $C_{(1)} \doteq 0$ in (2.10a) is implied by (3.3a). Similarly, once $C_{(1)}$, $C_{(2)}$ and $C_{(3)}$ vanish, then (3.3f) - (3.3h) imply the vanishing $C_{(4)} \doteq 0$, when linear supersymmetry is required.

We next give some outlines of the invariance check of our total action. As is always the case in field theory, we have to keep in mind that we can not use any field equations for this invariance check. For example, we can *not* assume the constancy of $B_\mu$. Therefore the variation of $\mathcal{L}_0$ under supersymmetry gives

$$\begin{aligned}\delta_Q\mathcal{L}_0 &= \tfrac{1}{4}(\overline{\epsilon}N\gamma^\nu\gamma^{\rho\sigma}\gamma^\mu\lambda^I)F_{\rho\sigma}{}^I(\partial_\mu B_\nu) - f^{IJK}(\overline{\epsilon}N\gamma_\mu\lambda^I)(\overline{\lambda}^J\gamma^\mu\lambda^K) \\ &= \tfrac{1}{4}(\overline{\epsilon}N\gamma^\nu\gamma^{\rho\sigma}\gamma^\mu\lambda^I)F_{\rho\sigma}{}^I C_{(2)\mu\nu} \\ &\quad - f^{IJK}\bigl[(\overline{\epsilon}N\gamma^\mu P\lambda^K)(\gamma_\mu)^{\alpha\beta} + 2(\overline{\epsilon}N)^{(\alpha|}(P\lambda^K)^{|\beta)}\bigr]C_{(1)\beta J}C_{(1)\alpha I} \ . \end{aligned} \tag{3.4}$$

Interestingly, the $\lambda^3$-term turns out to be proportional to the bilinear of the component $C_{(1)\alpha I} \equiv (N\lambda^I)_\alpha$, and therefore it can be cancelled by the variation of $\Xi^{(1),(1)}$ in (3.1f) *via* $\delta_Q\mathcal{L}_\Xi$. Relevantly, the important identities used are

$$f^{IJK}(\overline{\epsilon}N\gamma_\mu P\lambda^I)(\overline{\lambda}^J N\gamma^\mu P\lambda^K) \equiv 0 \ , \quad f^{IJK}(\overline{\epsilon}N\gamma_\mu N\lambda^I)(\overline{\lambda}^J P\gamma^\mu N\lambda^K) \equiv 0 \ , \tag{3.5a}$$

$$f^{IJK}(\overline{\epsilon}N\gamma_\mu P\lambda^I)(\overline{\lambda}^J P\gamma^\mu P\lambda^K) \equiv 0 \ , \quad f^{IJK}(\overline{\epsilon}N\gamma_\mu P\lambda^I)(\overline{\lambda}^J N\gamma^\mu N\lambda^K) \equiv 0 \ . \tag{3.5b}$$

Eq. (3.5a) can be understood as the 10D chiral Fierz identities, while (3.5b) can be easily obtained from $N\beta P = -NP = +NP \equiv 0$ *via* (2.5c). Due to these identities, there is only one sort of terms left over among these $\lambda$-trilinear terms, namely two of the three $\lambda$'s have the negative chirality, *i.e.*, either $N\lambda$ or $\overline{\lambda}P$, yielding the bilinear form of $C_{(1)}$.

The $C_{(2)}$-term in (3.4) can be cancelled by the variation of $\Lambda_{(2)}^{\mu\nu}$ in (3.1d) in $\delta_Q \mathcal{L}_\Lambda$. Thus all the terms generated in $\delta_Q \mathcal{L}_0$ are already cancelled by the like terms in $\delta_Q \mathcal{L}_\Lambda$ and $\delta_Q \mathcal{L}_\Xi$. The only remaining terms are confined within $\delta_Q \mathcal{L}_\Lambda$ and $\delta_Q \mathcal{L}_\Xi$. However, there is no new term from the former, because $\delta_Q C_{(2)\mu\nu} \equiv 0$. Now the remaining ones are only from $\delta_Q \Xi^{\mu_1\cdots\mu_r(i),\,\nu_1\cdots\nu_s(j)}$, where the first term in (3.1f) has been already used to cancel the $\lambda^3$-term in $\delta_Q \mathcal{L}_0$, so only the $M$-containing terms in (3.1e) and (3.1f) have new contributions. These terms are easily shown to cancel each other, because of our general formula using the $M$'s in (3.1e):

$$\delta_Q \mathcal{L}_\Xi|_M = +\tfrac{1}{2}\sum_{r=0}^{\infty}\sum_{s=0}^{\infty}\Big[-2\sum_{t=0}^{\infty}\Xi^{\mu_1\cdots\mu_r(i),\,\rho_1\cdots\rho_t(k)}M_{\rho_1\cdots\rho_t(k)}{}^{\nu_1\cdots\nu_s(j)}\Big]C_{\nu_1\cdots\nu_s(j)}C_{\mu_1\cdots\mu_r(i)}$$
$$+\sum_{r=0}^{\infty}\sum_{s=0}^{\infty}\Xi^{\mu_1\cdots\mu_r(i),\,\nu_1\cdots\nu_s(j)}\Big[\sum_{t=0}^{\infty}M_{\nu_1\cdots\nu_s(j)}{}^{\rho_1\cdots\rho_t(k)}C_{\rho_1\cdots\rho_t(k)}\Big]C_{\mu_1\cdots\mu_r(i)} = 0 \ , \quad (3.6)$$

where the symbol $|_M$ in the l.h.s. stands only for the $M$-containing terms, skipping the first term in (3.1f). Needless to say, the Grassmann parities for all the *hidden* indices accompanying $(i), (j), \cdots$ have been also taken into account. This concludes the invariance of our total action for the lagrangian $\mathcal{L} \equiv \mathcal{L}_0 + \mathcal{L}_\Lambda + \mathcal{L}_\Xi$ under supersymmetry.

We next address ourselves to the consistency of the field equation (2.12) with supersymmetry. The variation of (2.12b) under supersymmetry is

$$\delta_Q(\partial_\mu \Lambda_{(2)}^{\mu\nu}) = -\tfrac{1}{4}(\overline{\overline{\epsilon}}\gamma^\nu\gamma^{\rho\sigma}\gamma^\mu D_\mu \widehat{\lambda}^I)F_{\rho\sigma} - \tfrac{1}{4}(\overline{\overline{\epsilon}}\gamma^\nu\gamma^{\rho\sigma}\gamma^\mu \widehat{\lambda}^I)D_\mu F_{\rho\sigma}$$
$$\doteq +\tfrac{1}{2}f^{IJK}(\overline{\overline{\epsilon}}\gamma^\nu\gamma^\rho\widehat{\lambda}^I)(\overline{\widehat{\lambda}}{}^J\gamma_\rho\widehat{\lambda}^K) \ , \quad (3.7)$$

with $\widehat{\lambda} \equiv P\lambda$, $\overline{\widehat{\lambda}} \equiv \overline{\lambda}N$, $\overline{\overline{\epsilon}} \equiv \overline{\epsilon}N$ upon the usage of (2.10), (2.13) and (2.14), as well as the Bianchi identity $D_{[\mu}F_{\nu\rho]}{}^I \equiv 0$. Now using $\overline{\overline{\epsilon}} = -\overline{\overline{\epsilon}}\beta$, and the identity

$$\beta\gamma^\nu = -\gamma^\nu\beta + 2B^\nu \ , \quad (3.8)$$

we see that the last line of (3.7) is equal to $(-1)$ times the first side plus a term $f^{IJK}B^\nu(\overline{\overline{\epsilon}}\gamma_\mu\widehat{\lambda}^I)(\overline{\widehat{\lambda}}{}^J\gamma^\mu\widehat{\lambda}^K) \equiv 0$ upon (3.5a). Therefore (3.7) is to be zero by itself.

As some careful readers may have noticed, we have to consider the question of finiteness of the total action, since we have an infinite sum of terms in the lagrangian [14]. In our formulation, however, this poses no problem, because any higher-order terms with $C_{\mu_1\cdots\mu_r(i)}$ at $r \geq 1$ vanish, which are nothing else than higher covariant derivatives acting on the field equations (constraints) (2.10). Moreover, there will arise no new conditions or field equations, because they will be always necessary conditions of the field equations (2.10). Therefore, there arises no problem with the finiteness of the total action.

## 4. Closure of Supersymmetry

The only remaining important task is the confirmation of the on-shell closure of supersymmetry algebra (3.1): $[\delta_Q(\epsilon_1), \delta_Q(\epsilon_2)] \equiv [\delta_1, \delta_2] = \delta_P(\xi^\mu)$ with the translation parameter $\xi^\mu \equiv +(\overline{\epsilon}_1 N\gamma^\mu P\epsilon_2)$. This is not too difficult, because for the on-shell closure check, we can use the field equations (2.10), (2.12) - (2.14). The closure on $A_\mu{}^I$ goes very easily with no particular comment here. As for the closure $[\delta_1, \delta_2]\lambda_\alpha{}^I$, we can project out only into the 'positive chirality' component $[\delta_1, \delta_2]P\lambda_\alpha{}^I$. We also need as usual Fierzing of fermionic bilinears in 11D:

$$(\overline{\chi}\psi)(\overline{\lambda}\omega) \equiv -\tfrac{1}{32} \sum_{n=0}^{5} \frac{(-1)^{n(n-1)/2}}{n!} (\overline{\chi}\gamma_{[n]}\omega)(\overline{\lambda}\gamma^{[n]}\psi) \quad , \tag{4.1}$$

as well as frequent usage of the gaugino field equation (2.14). After these, we can easily find out that only $\xi^\mu$-terms are left over, because, e.g., the factor $(\overline{\epsilon}N\gamma_{[5]}P\epsilon_2)$ vanish upon $\gamma$-algebra and the gaugino field equation, while the terms with the factor $(\overline{\epsilon}_1 N\gamma_{\mu\nu}P\epsilon_2)$ are shown to be proportional to the term with $\xi^\mu$. Typical relationships used are (2.5) and

$$\gamma_\mu P = N\gamma_\mu + B_\mu \quad , \qquad P\gamma_\mu = \gamma_\mu N + B_\mu \quad . \tag{4.2}$$

Even though it looks trivial at first sight, the closure on $\varphi$ provides a good consistency check. First, (3.1c) implies $[\delta_1, \delta_2]\varphi \equiv 0$, suggesting that $\varphi$ is independent of the coordinates. However, we also know that $\varphi$ has some on-shell non-trivial dependence on the coordinate via (2.8). We see that our system cleverly avoids this superficial contradiction by the algebra:

$$\begin{aligned}{}[\delta_1, \delta_2]\varphi = \xi^\mu \partial_\mu \varphi &= (\overline{\epsilon}_1 N\gamma^\mu P\epsilon_2)\partial_\mu \varphi \\ &= (\overline{\epsilon}_1 N\gamma^\mu P\epsilon_2)B_\mu\sqrt{(\partial_\nu\varphi)^2} = (\overline{\epsilon}_1 N\beta P\epsilon_2)\sqrt{(\partial_\nu\varphi)^2} \equiv 0 \quad , \end{aligned} \tag{4.3}$$

based on the previously-mentioned identity $N\beta P = -NP = +NP \equiv 0$.

The closure on $\Lambda_{(2)}^{\mu\nu}$ needs a particular care, because it generates some 'extra' symmetry we have not yet mentioned:

$$\delta_{\mathrm{E}}\Lambda_{(2)}^{\mu\nu} = \xi^\nu \partial_\nu \Lambda_{(2)}^{\mu\nu} + \partial_\tau \Omega^{\nu[\mu\tau]} + \xi^\nu X^\mu + B^\nu Y^\mu \quad , \tag{4.4}$$

where $\Omega^{\nu[\mu\tau]}$, $X^\mu$ and $Y^\mu$ are arbitrary field-dependent functions, while $\xi^\mu \equiv (\overline{\epsilon}_1 N\gamma^\mu P\epsilon_2)$. Note that the first term is nothing else than the translation operation itself. In other words, for this field $\Lambda_{(2)}^{\mu\nu}$, the translation itself can be interpreted as an 'extra' symmetry. In fact, the identity $\xi^\mu \partial_\mu \varphi \equiv 0$ in (4.3) implies that $\xi^\mu \partial_\mu B_\nu \equiv 0$ which is not a field equation, but just an algebraic identity. It then follows that

$$\begin{aligned} \delta'_{\mathrm{E}} \mathcal{L}_\Lambda &= (\delta'_{\mathrm{E}}\Lambda_{(2)}^{\mu\nu})\partial_\mu B_\nu = (\xi^\rho \partial_\rho \Lambda_{(2)}^{\mu\nu})\partial_\mu B_\nu \\ &= -\xi^\rho \Lambda_{(2)}^{\mu\nu} \partial_\rho \partial_\mu B_\nu = -\Lambda_{(2)}^{\mu\nu} \partial_\mu(\xi^\rho \partial_\rho B_\nu) \equiv 0 \quad , \end{aligned} \tag{4.5}$$

up to a total divergence, where the symbol $\delta'_{\mathrm{E}}$ stands for the first term in (4.4) for the translation. We can easily confirm the validity of other extra symmetries in (4.4), due to the identity $B_\mu^2 \equiv 1$ with its necessary condition $B^\mu \partial_\nu B_\mu \equiv 0$. After all, by explicit computations of $[\delta_1, \delta_2]\Lambda_{(2)}^{\mu\nu}$, we can confirm that all the terms generated can be interpreted as one of these extra symmetries (4.4), or vanishing terms upon the usage of field equations.

The only remaining closure is on the fields $\Xi^{\mu_1\cdots\mu_r(i),\nu_1\cdots\nu_s(j)}$. Note first that our $\Xi$-multiplier fields differ slightly from those used in [13], in the sense that there are no 'kinetic'-type bilinear terms for $C$'s, except for those accompanying the $\Xi$'s. Therefore it is difficult to get the Siegel-type invariance [13] at the lagrangian level that can gauge away all the $\Xi$'s. However, this poses no problem at all, because these multipliers $\Xi$ are still decoupled from all the field equations. This can be understood from the general feature in field theory that the system of field equations has a larger covariance symmetry than the lagrangian invariance. This feature has been known for some time in different contexts, such as the $U$-duality [17] in $N=8$ supergravity in 4D [18], in which the $E_{7(+7)}$ symmetry is realized as on-shell symmetry among field strengths that is not manifest at the lagrangian level. In other words, since the $\Xi$-fields disappear from all the field equations, there are larger symmetries that can gauge away all the $\Xi$-fields entirely from the system of field equations. In terms of path-integral language, all the $\Xi^{\mu_1\cdots\mu_r(i),\nu_1\cdots\nu_s(j)}$-fields are playing a role of multiplier fields that can be completely integrated out, yielding the functional $\delta$-function $\delta(C_{\mu_1\cdots\mu_r(i)}C_{\nu_1\cdots\nu_s(j)})$. Considering these points, we conclude that the *on-shell* closure of supersymmetry on the $\Xi$-fields poses no problem.

Before concluding this section, we mention the commutators involving the $SO(10,1)$ Lorentz transformation $\delta_L(\alpha^{\rho\sigma})$ in 11D with the parameter $\alpha^{\rho\sigma}$. Since Lorentz symmetry is broken at the field equation level, a non-trivial question is about the validity of commutator algebra with $\delta_L(\alpha^{\rho\sigma})$. The main point here is that $B_\mu$ transforms covariantly, while the operators $\beta$, $P$ and $N$ transform just like a Dirac matrix with no index:

$$\delta_L B_\mu = \alpha_\mu{}^\nu B_\nu \ , \quad \delta_L \beta = \alpha_\mu{}^\nu B_\nu \gamma^\mu = [\tfrac{1}{4}\alpha^{\rho\sigma}\gamma_{\rho\sigma}, \beta] \ ,$$
$$\delta_L P = [\tfrac{1}{4}\alpha^{\rho\sigma}\gamma_{\rho\sigma}, P] \ , \quad \delta_L N = [\tfrac{1}{4}\alpha^{\rho\sigma}\gamma_{\rho\sigma}, N] \ , \tag{4.6}$$

due to invariance of $\gamma_{\rho\sigma}$: $\delta_L \gamma_{\rho\sigma} = 0$. Using these relations, it is straightforward to conclude that

$$[\delta_L(\alpha^{\rho\sigma}), \delta_Q(\epsilon)] = \delta_Q\bigl(\tfrac{1}{4}\alpha^{\rho\sigma}\gamma_{\rho\sigma}\epsilon\bigr) \equiv \delta_Q(\epsilon') \ , \tag{4.7}$$

on all the fundamental fields with the new parameter $\epsilon' \equiv (1/4)\alpha^{\rho\sigma}\gamma_{\rho\sigma}\epsilon$ for supersymmetry, as desired. Additionally, using also the crucial relationships in (4.6) for $\mathcal{L}_\Lambda$ and $\mathcal{L}_\Xi$, we encounter no problem with confirming the invariance of our total lagrangian $\mathcal{L}$ under 11D Lorentz transformations.

## 5. Supermembrane Action

Once we have established the component formulation of Yang-Mills vector multiplet in 11D, our next natural question is the existence of supermembrane action [19]. Even though we do not yet have the supergravity sector in 11D, we are still able to construct such an action on the Yang-Mills backgrounds. Our total action has no Wess-Zumino-Novikov-Witten (WZNW) term for conventional supermembranes [19], but instead it has a topological Chern-Simons-Hopf term.[6]

Our total action $I$ is composed of $I_0$ and $I_{\text{CSH}}$:

$$I \equiv I_0 + I_{\text{CSH}} \ ,$$
$$I_0 \equiv \int d^3\sigma \left[ \tfrac{1}{2}\sqrt{-g} g^{ij}\eta_{ab}\Pi_i{}^a\Pi_j{}^b - \tfrac{1}{2}\sqrt{-g} \right] \ ,$$
$$I_{\text{CSH}} \equiv \int d^3\sigma \left[ k\, \epsilon^{ijk}\Pi_i{}^C\Pi_j{}^B\Pi_k{}^A(F_{AB}{}^I A_C{}^I - \tfrac{1}{3}f^{IJK}A_A{}^I A_B{}^J A_C{}^K) \right] \ . \quad (5.1)$$

Only in this section, we use the superspace local Lorentz indices $A \equiv (a,\alpha),\ B \equiv (b,\beta),\ \cdots$ for bosonic $a,\ b,\ \cdots$ and fermionic ones $\alpha,\ \beta,\ \cdots$. As usual, $\Pi_i{}^A \equiv (\partial_i Z^M)E_M{}^A$ is the superspace pull-backs. The $k$ in $I_{\text{CSH}}$ is a real constant. Compared with the conventional supermembrane theory [19], the lack of the antisymmetric superfield $B_{AB}$ invalidates the usual WZNW term, while the presence of the Yang-Mills background necessitates the topological Chern-Simons-Hopf term $I_{\text{CSH}}$.

From the component result (3.1) and the supersymmetry closure, we can see the relevant non-vanishing background superspace constraints are

$$T_{\alpha\beta}{}^c = +(N\gamma^c P)_{\alpha\beta} \ , \qquad P \equiv \tfrac{1}{2}(I+\beta) \ , \qquad N \equiv \tfrac{1}{2}(I-\beta) \ ,$$
$$F_{\alpha b}{}^I = +(N\gamma_b)_\alpha{}^\beta \lambda_\beta{}^I \equiv +(N\gamma_b \lambda^I)_\alpha \ , \qquad \beta \equiv B_a \gamma^a \ ,$$
$$\nabla_\alpha \lambda_\beta{}^I = -\tfrac{1}{4}(N\gamma^{ab})_{\alpha\beta} F_{ab}{}^I \ , \qquad B_a \equiv \tfrac{\nabla_a \varphi}{\sqrt{(\nabla_b\varphi)^2}} \ ,$$
$$\nabla_\alpha \varphi = 0 \ , \qquad \nabla_\alpha \Lambda^{ab}_{(2)} = \tfrac{1}{4}(N\gamma^b \gamma^{cd}\gamma^a \lambda^I)_{\alpha\beta} F_{cd}{}^I \ , \qquad C_{(i)} \doteq 0 \ , \quad (5.2)$$

where $\nabla_\alpha \Xi$ is skipped because of its lengthiness, but it is easily obtained from (3.1e) and (3.1f). The component $T_{\alpha\beta}{}^c$ is symmetric in $\alpha \leftrightarrow \beta$, as is easily confirmed. These constraints satisfy the usual $F$-Bianchi identities. As is well-known, superspace formulation in 11D is essentially *on-shell*, so that constraints such as $C_{(i)} \doteq 0$ are easily imposed. Our action in (3.1) has similarity to the model in [20] in the sense that a Chern-Simons-Hopf form is used in addition to the $\sigma$-model kinetic term, but has also difference that the 11D superspace is used here as the target space.

We can now show that our total action $I$ is invariant under the fermionic $\kappa$-symmetry

$$\delta_\kappa E^\alpha = (N\kappa)^\alpha \equiv \tfrac{1}{2}(I-\beta)^{\alpha\beta}\kappa_\beta \ , \qquad \delta_\kappa E^a = 0 \ , \qquad \delta_\kappa A_A{}^I = (\delta_\kappa E^\beta)E_\beta A_A{}^I \ , \quad (5.3)$$

where $E_A \equiv E_A{}^M \partial_M$ [21], and $\delta_\kappa E^A \equiv (\delta_\kappa Z^M)E_M{}^A$ [19]. Compared with the conventional supermembrane theory [19], the projection operator $(I+\Gamma)$ in [19] is replaced by the projection operator $N \equiv (I-\beta)/2$. This is the main similarity as well as the difference at the same time. The background and the $\sigma$-model quantities also satisfy the constraints

$$(N)_{\alpha\beta}\Pi_i{}^\beta = 0 \ , \tag{5.4a}$$
$$\Pi_i{}^a B_a = 0 \ . \tag{5.4b}$$

---

[6]We use this terminology, because it is a mixture of Hopf terms with pull-backs and Chern-Simons terms.

The constraint (5.4b) is similar to those constraints in supergravity in 12D or higher [4], because the space-like unit vector is used, like the null-vectors in the latter.

The confirmation of $\delta_\kappa I = 0$ goes as in the 1-st order formulation of supermembrane theory [19], in which we do not vary the 3D metric $g_{ij}$, satisfying the algebraic embedding condition [19]

$$g_{ij} \doteq \eta_{ab} \Pi_i{}^a \Pi_j{}^b \quad , \tag{5.5}$$

as its field equation. Other relevant useful relations are such as

$$\delta_\kappa A_i{}^I = D_i[\, (\delta_\kappa E^A) A_A{}^I \,] + (\delta_\kappa E^B) \Pi_i{}^A F_{AB}{}^I \quad ,$$
$$\delta_\kappa F_{ij}{}^I = D_i(\delta_\kappa A_j{}^I) - D_j(\delta_\kappa A_i{}^I) \quad , \tag{5.6}$$

where $A_i{}^I \equiv \Pi_i{}^A A_A{}^I$, $F_{ij}{}^I \equiv \Pi_j{}^B \Pi_i{}^A F_{AB}{}^I$, and the derivative $D_i$ is Yang-Mills covariant with the composite connection $\Pi_i{}^A A_A{}^I$. Interestingly, the 'composite' Bianchi identity $D_{[i} F_{ij]}{}^I \equiv 0$ is easily confirmed. With these useful formulae, the invariance check of the total action is straightforward, whose further details are skipped in this paper.

## 6. Concluding Remarks

We have in this Letter presented a Lorentz invariant lagrangian formulation of a Yang-Mills vector multiplet in 11D, which had been regarded as 'non-existent', based on the well-known no-go theorem [2]. The key ingredient of our formulation is the introduction of the space-like 'harmonic' unit vector [12] $B_\mu \equiv (\partial_\mu \varphi)/\sqrt{(\partial_\nu \varphi)^2}$ in terms of the scalar field $\varphi$, avoiding the 'constant' vector that explicitly break Lorentz invariance of the lagrangian. The 'on-shell' constancy of $B_\mu$ is implied by the field equation of the multiplier field $\Lambda^{\mu\nu}_{(2)}$ and the Siegel's multiplier $\Xi^{(2),(2)}$ [13]. In other words, the 'constancy' of $B_\mu$ and therefore the breaking of Lorentz covariance occurs only at the field equation level like spontaneous breaking, keeping the Lorentz invariance of the lagrangian intact. This mechanism is also similar to the constant field strengths breaking Lorentz covariance in non-commutative geometry formulation [6] in string theory. We have seen that the invariance of the lagrangian is guaranteed by explicit computations, while the on-shell closure of supersymmetry is also confirmed by the use of field equations. We have seen that the infinitely many terms with the multiplier field $\Xi$ are needed, but effectively they are to be terminated as a finite sum in order for the total action to be finite.

Based on the component result, we have also developed a supermembrane action in the Green-Schwarz $\sigma$-model formulation which is invariant under the fermionic $\kappa$-symmetry. Interestingly, the usual WZNW term is replaced by a Chern-Simons-Hopf term with supersymmetric Yang-Mills background. The existence of such a supermembrane action also supports the validity of our Yang-Mills formulation itself in 11D.

The physical degrees of freedom in our system have no contradiction with the common wisdom with supersymmetry. This is because of the constraints (2.10a) and (2.10c) that eliminate the unwanted redundant degrees of freedom from 9 for $A_\mu$ plus 16 for $\lambda$ into $8 + 8$ as in the 10D case.

In our formulation, we have also seen the existence of huge duality symmetries that can gauge away all the multiplier fields $\Xi^{\mu_1\cdots\mu_r(i),\nu_1\cdots\nu_s(j)}$ at the field equation level, which are larger than the lagrangian invariance. This is because these multipliers are completely decoupled from any field equations. This is similar to other duality symmetry, such as the $U$-duality [17] among the field strengths in $N=8$ supergravity in 4D [18]. This feature is suggestive of other duality symmetries inherent in the system that are possibly related to the general vacuum degeneracy in M-theory [9][10].

The recent development, associated with the $U(1)$ gauge fields on the D-brane worldvolumes described by supersymmetric Born-Infeld action [22][23][24], also strongly motivates our search of such vector multiplets in 11D and beyond, not to mention the search for a Born-Infeld action based on our system in 11D itself. Armed with these lowest order terms in the lagrangian, it is to be straightforward to fix supersymmetric Born-Infeld lagrangian [23][24] with higher-derivatives within 11D. Thus our result in this paper has made it possible to consider a supersymmetric Born-Infeld action within 11D, that had been impossible in the past.

Some readers may be wondering, if our formulation is completely equivalent to the conventional 10D supersymmetric Yang-Mills theory. However, we point out some possibilities, related to the constraint solutions in (2.10a) and (2.10d), that we have not mentioned. The argument we used there was that if $\psi_\alpha \psi_\beta = 0$ for a spinor $\psi_\alpha$ with *more than* 4 components, then the only solution is $\psi_\alpha = 0$. In fact, if $\psi_\alpha$ has $n$ components for $n \geq 4$, then $\psi_\alpha \psi_\beta = 0$ has in total $n(n-1)/2$ conditions. Since $n(n-1)/2 > n$ for $n \geq 4$, $\psi_\alpha \psi_\beta = 0$ is sufficient to conclude $\psi_\alpha = 0$. This argument is valid, only when we consider only *linear* supersymmetry. In fact, a good example is the usual prescription for the superfield equation $\Phi^2 = 0$ for a real scalar superfield $\Phi$ implying $\Phi = 0$ in 4D, which is equivalent to conclude $\chi_\alpha = 0$ from $\theta^\alpha \theta^\beta \chi_\beta \chi_\alpha = 0$ for a 4-components spinor $\chi_\alpha$. However, it has been also known in *non-linear* realization of supersymmetry [16], that the condition $\psi_\alpha \psi_\beta = 0$ can allow solutions for $\psi_\alpha$ with intrinsic nilpotent quantities, such as $\psi_\alpha = u_\alpha \zeta$, where $\zeta$ is nilpotent $\zeta^2 \equiv 0$ with a 'commuting' spinor $u_\alpha$. Such a realization breaks the usual linear supersymmetry. As a matter of fact, if such a nilpotent is allowed, then the spinor $\psi_\alpha = u_\alpha \zeta$ satisfies the usual Dirac equation $\gamma^\mu D_\mu u = 0$, while its energy-momentum tensor *vanishes* identically, breaking the usual boson-fermion matching in linear supersymmetry. Another important aspect is that the solutions in (2.10) are valid only *on-shell*, but have never been used in the invariance confirmation of the total action under supersymmetry. From these viewpoints, we see that our system is more complicated with non-linear supersymmetry than just a 'rewriting' of the conventional 10D supersymmetric Yang-Mills, or more non-trivial than just the latter in 11D in 'disguise'. To put it differently, the vacuum structure of our 11D system is more involved than the conventional 10D supersymmetric Yang-Mills theory. We expect this is also associated with the non-trivial vacuum structure of underlying M-theory [9][10], accommodating even non-linear realization of supersymmetries [16][25].

Our result in this paper might have some possible link with the formulation in ref. [26], in which the 10D superfield formulation with infinitely many constraint terms is 'oxidized' from 10D to 11D, by adding an extra coordinates equivalent to the infinite set of auxiliary fields. Even though the original motivation of the formulation in [26] is to maintain the manifest Lorentz invariance within 10D in terms of superfields, there may well be some close connection or duality between these two formulations, sharing the idea of 'oxidation' from 10D to 11D.

Another natural next step is to consider the possible coupling with supergravity in 11D. Actually, such a trial is not too far-fetched, considering the recent development of M-theory indicating a close relationship between supersymmetric Yang-Mills in 10D and supergravity in 11D [9][10]. Such a formulation may be not too difficult, once we have understood the main ingredient of the global supersymmetry, even though there might arise some subtlety about making the global Lorentz symmetry into a local one. Again the combination of space-like 'harmonic' vectors [12] and the Siegel's multipliers [13] are supposed to play crucial roles in such a formulation.

Compared with other formulations of supersymmetric Yang-Mills theories in 12D or higher dimensions with two time coordinates with null-vectors [3], our formulation has similarities as well as differences. The similarity is existence of the space-like vector $B_\mu$ [12] in our present paper in place of null-vectors in [3], while the difference is that the space-like vector has the unit magnitude $B_\mu^2 = 1$, instead of zero for null-vectors in [3]. Another similarity is that a scalar field is used to define $B_\mu$, like those null-vectors in higher-dimensions defined in terms of scalars [3].

The validity of our supersymmetric Yang-Mills multiplet in 11D suggests similar possibilities in $D \geq 12$ with only one time coordinate in a parallel fashion, but distinct from those formulations with two time coordinates [3][4]. To put it differently, there seems to be no explicit limit for supersymmetric theories in general higher dimensions. In fact, some similar symptom has been recently reported for non-linear formulations of supersymmetries in $^\forall D$ and $^\forall N$ [25]. This series of works in infinitely many space-time dimensions indicate that more new formulations of global supersymmetries in arbitrarily higher-dimensions are to be discovered in the future.

We are grateful to J. Gates, Jr. and M. Roček for helpful discussions.